\documentclass[aip,sd,showkeys,superscriptaddress,reprint]{revtex4-2}

\usepackage{amsfonts,amsmath,amsthm,amssymb}
\usepackage{graphicx,epsfig,pstricks,pst-node}
\usepackage{setspace}
\usepackage{times}
\begin{document}

\title{Neutral species facilitate coexistence among cyclically competing species under birth and death processes}

\author{Yikang Lu}
\affiliation{School of Statistics and Mathematics, Yunnan University of Finance and Economics, Kunming, Yunnan 650221, China}

\author{Wenhao She}
%\thanks{Corresponding authors}
%\email{18955034301@163.com}
\affiliation{School of Statistics and Mathematics, Yunnan University of Finance and Economics, Kunming, Yunnan 650221, China}

\author{Xiaofang Duan}
%\thanks{Corresponding authors}
%\email{duanxf427@163.com}
\affiliation{School of Mathematics and Statistics, Xidian University, Xi’an, Shaanxi 710126, China}

\author{Junpyo Park}
%\thanks{Corresponding authors}
%\email{junpyopark@khu.ac.kr}
\affiliation{Department of Applied Mathematics, 
	Colleage of Applied Sciences,
	Kyung Hee University, 
	Yongin 17104, Republic of Korea}
\affiliation{G-LAMP NEXUS Institute, 
	Kyung Hee University, 
	Yongin 17104, Republic of Korea}

\date{\today}
	
\begin{abstract}
Natural birth and death are fundamental mechanisms of population dynamics in ecosystems and have played pivotal roles in shaping population dynamics. Nevertheless, in studies of cyclic competition systems governed by the rock-paper-scissors (RPS) game, these mechanisms have often been ignored in analyses of biodiversity.
On the other hand, given the prevalence and profound impact on biodiversity, understanding how higher-order interactions (HOIs) can affect biodiversity is one of the most challenging issues, and thus HOIs have been continuously studied for their effects on biodiversity in systems of cyclic competing populations, with a focus on neutral species.
However, in real ecosystems, species can evolve and die naturally or be preyed upon by predators, whereas previous studies have considered only classic reaction rules among three species with a neutral, nonparticipant species. 
To identify how neutral species can affect the biodiversity of the RPS system when species' natural birth and death are assumed, we consider a model of neutral species in higher-order interactions within the spatial RPS system, assuming birth-and-death processes. Extensive simulations show that when neutral species interfere positively, they dominate the available space, thereby reducing the proportion of other species. Conversely, when the interference is harmful, the density of competing species increases. In addition, unlike traditional RPS dynamics, biodiversity can be effectively maintained even in high-mobility regimes. Our study reaffirms the critical role of neutral species in preserving biodiversity.
\end{abstract}
		

\maketitle

\begin{quotation}
Understanding the mechanisms that maintain biodiversity remains a persistent challenge, and various methodologies are used to elucidate this issue. Typically, approaches focus on pairwise interactions, in which two species interact directly. However, this approach has limitations for elucidating the complex characteristics of ecosystems or for understanding interactions among multiple organisms; consequently, methods that incorporate higher-order interactions are being proposed. In particular, the intervention of neutral species has been reported to affect biodiversity in spatial systems of the rock-paper-scissors (RPS) game, a representative evolutionary game model used to describe ecosystems. As the number of neutral species that do not participate in actual competition increases, the natural birth and death of populations and their impact on existing ecosystems are obscured. We explore the effect of neutral species, which do not directly compete with existing populations, on biodiversity, considering the fundamental characteristics of individual birth and death. The presence of neutral species has been shown to have both positive and negative effects on ecosystems, and, unlike previous RPS systems, to have the potential to sustain biodiversity. Through our work, we can reconfirm the critical role of neutral species in conserving biodiversity.
\end{quotation}

\section{\label{sec:intro}Introduction}

The mechanisms by which biodiversity is maintained have persistently challenged and perplexed academics in the ecological and biological sciences.~\cite{Cardinale2002Species,Schmidt2011Persistence} 
To address such a complex issue, scientists have usually focused on pairwise interactions, in which two species interact directly.~\cite{szolnoki2024When,szolnoki2016zealots,Reichenbach2007Mobility,Brophy2017Biodiversity,Park2017Emergence,szolnoki2004Phase} 
Subsequently, they proposed and investigated the effects of higher-order interactions (HOIs) on biodiversity. 
Higher-order interactions occur when the interaction between two species is influenced by additional species or factors.~\cite{Bairey2016High} 
Over the past decades, extensive research has focused on these higher-order interactions and their effects on the maintenance of biodiversity.~\cite{szolnoki2020strategy,szolnoki2020pattern,Majhi2022Dynamics,Majhi2017Dynamics}

In studies on biodiversity, the rock-paper-scissors game (RPS) has been a fundamental tool for examining non-hierarchical, cyclic competitive interactions, elucidating diverse mechanisms in ecosystems, including mating strategies among three side-blotched lizard populations in California~\cite{sinervo1996rock} and the evolution of microbial populations.~\cite{kirkup2004antibiotic, neumann2010evolutionary, nahum2011evolution}
As the study in theoretical frameworks is identified to support experimental studies that highlight the importance of mobility for biodiversity,~\cite{Reichenbach2007Mobility} studies of cyclic competition have been extensively examined by considering various factors that affect species biodiversity from the perspective of pairwise interactions, e.g., the double-edged sword effect of cooperation on promoting coexistence,~\cite{duan2024does} species characteristics,~\cite{lu2022Effects,Souza2017Apex} anthropogenic influences and the consideration of a wild refuge,~\cite{lu2022Enhancing} and the beneficial role of the weakest species in promoting coexistence.~\cite{Avelino2019Predominance,Liao2020Survival}
Even if previous studies have focused on pairwise interactions, higher-order interactions have recently gained attention and have become one of the interesting issues in this field.~\cite{Liao2020Survival,lu2024preUnderstanding,Chatterjee2022Controlling,Gibbs2022Coexistence,Lu2025Enhancement}
To be concrete, the authors in Ref.~[\onlinecite{Chatterjee2022Controlling}] introduced a novel framework for understanding how higher-order interactions shape the evolution of social phenotypes. Their findings indicate that a perturbed interaction network can establish coexistence equilibria among competing species, highlighting an inverse relationship between the ratio of unperturbed to perturbed species and the magnitude of perturbations.
In other scenarios involving higher-order interactions with neutral species, the increase in mobility can promote species coexistence.~\cite{lu2024preUnderstanding}
In addition, following the work, the intensity of competition among different species on spatially embedded hyper-lattices has been shown to promote species coexistence through higher-order forms of competition.~\cite{Lu2025Enhancement}

As the importance of higher-order interactions increases for elucidating complex phenomena in ecosystems, it is becoming necessary to consider them. In addition, within the framework of HOIs, the significant role of neutral species has been revealed as one of the beneficial factors in promoting biodiversity and maintaining coexistence{~\cite{2007SzabSegregation, 2019BazeiaInvasion,2018SzolnokiEvolutionary}.}
{Relevant works indicate that third-party species, which can be considered neutral in real ecosystems, are able to maintain system stability by modulating competitive effects. For example, the presence of a third-party tree can mitigate the direct competitive effects of neighboring trees on focal trees, thereby sustaining forest species diversity~\cite{2021LiBeyond}.}
For species survival, especially coexistence, previous studies have shown that sufficient space (also regarded as carrying capacity) is required.~\cite{PDHL:2013}
In spatial RPS systems, such spaces are generated by predation under cyclic competition, in the classic manner, and additional interactions, e.g., intraspecific competition, can substantially produce more empty sites, which eventually leads to the coexistence of species beyond the effect of mobility.~\cite{Park2017Emergence,Park:2018, PJ:2021,PJ:2023}
Another way to enhance carrying capacity in spatially extended systems is natural death, a fundamental feature of populations, which has been shown in experimental studies to play a pivotal role in driving changes in population dynamics.~\cite{Xiao:2000,LiChen:2009,Holt:2013,Then:2014,McCoy:2008} 
Because of this, the effects of natural birth and death have been widely discussed theoretically in the field of evolutionary games, particularly in terms of evolutionarily stable strategy and collective behavior.~\cite{Knebel2015coupleddeath,Yu2016QBD,West2020fixation}

Nevertheless, spatial dynamics become increasingly complex due to the presence of neutral species that do not participate in actual competition. While these neutral species are known to promote biodiversity and coexistence among cyclically competing populations, they also relatively reduce spatial capacity. Therefore, even when natural birth and death of individuals are considered, the impact of HOIs by neutral species on biodiversity becomes ambiguous. In this context, this paper aims to analyze the impact of neutral species on biodiversity in the RPS system within the framework of HOIs that assume species' nature.

Contrary to prior studies that have typically discussed the competing species,~\cite{Schmidt2011Persistence,Brophy2017Biodiversity} our model assumes that neutral species do not engage in direct competition, {they generate higher-order interactions by altering the strength of interactions among other species and modulating predation rates, thereby affecting the local environment and spatial structure.} They occupy available resource sites within the system, represented as empty sites in the RPS game. Neutral species can occupy these empty sites and subsequently face natural mortality.
Natural birth is assumed for all species and is realized through reproduction in the fundamental reaction rules of the RPS game.
The presence of neutral species disrupts the ecosystem, thereby influencing competition rates among the other species in the RPS game.~\cite{He2012Coexistence} A sensitivity coefficient modulates these rates: when the coefficient is greater than $0$, the neutral species enhances competition among established species; otherwise, it suppresses competition.

Through extensive simulations, we found that the total density of species, excluding neutral species, increases with the sensitivity coefficient: when the coefficient is less than zero, the neutral species $D$ dominates the ecosystem, compressing the proportions of other species and hindering their competitive interactions, thereby paradoxically supporting the maintenance of species diversity. 
{
Under these conditions, neutral species exert a similar effect at different mobility rates by impeding predatory interactions among competing species, thus preventing extinction driven by predation.} Furthermore, when measuring the extinction rate, we observed that it reaches zero at high sensitivity values, further reinforcing the idea that higher sensitivity coefficients favor species coexistence. 
Notably, under high-mobility conditions, the extinction probability remains zero. These findings align with our previous results shown in the patch diagram.
On the other hand, increasing species' reproduction rates to several times their standard values fundamentally alters the spatial patterns observed under standard conditions. High reproduction rates reduce the density of the neutral species within the system and, under asymmetric mortality between competitive and neutral species, give rise to spiral-wave patterns that support the coexistence of all species. Asymmetric mortality also reshapes extinction dynamics across intervention intensities and mobility levels. We find that maintaining the persistence of the neutral species more effectively promotes coexistence among the competitive species and mitigates the high extinction risk induced by strong mobility under low intervention intensity.

This paper is organized as follows. In Sec.~\ref{sec:model}, we provide a detailed description of our model, focusing on rock-paper-scissors dynamics with neutral species participating in the birth and death processes.
In Sec.~\ref{sec:result}, we present results on the characteristics of pattern formation, biodiversity, and extinction rates influenced by neutral species. We conclude with a summary of our findings in Sec.~\ref{sec:conc}.

\section{\label{sec:model}Model}

Consider four distinct species $A$, $B$, $C$, $D$, and empty sites $\varnothing$ on a square lattice of size $N = L \times L$.
Among four species, three species $A$, $B$, and $C$ are governed by a rock-paper-scissors (RPS) game and interact with neighboring sites on a network according to a set of rules:
\begin{eqnarray}
	&& AB \xrightarrow{~p~} A \varnothing, \quad 
	BC \xrightarrow{~p~} B \varnothing, \quad 
	CA \xrightarrow{~p~} C \varnothing, \label{eq1}\\
	&& A\varnothing \xrightarrow{~q~} AA, \quad 
	B\varnothing \xrightarrow{~q~} BB, \quad 
	C\varnothing \xrightarrow{~q~} CC, \label{eq2}\\
	&& A\square \xrightarrow{~\varepsilon~} \square A, \quad
	B\square \xrightarrow{~\varepsilon~} \square B, \quad
	C\square \xrightarrow{~\varepsilon~} \square C, 	\label{eq3}
\end{eqnarray}
where $\square$ denotes one of four species or an empty site.
Note that the neutral species $D$ does not participate in the competition of RPS game, but can change its position.
Eq.~(\ref{eq1}) describes the cyclically competitive interactions among species $A$, $B$, and $C$: species $A$ outcompetes species $B$ at rate $p$, leaving an empty site, and species $B$ outcompetes species $C$. 
In Eq.~(\ref{eq2}), the reproduction process of the species is described, in which each species disperses its offspring to neighboring empty sites at rate $q$.
Since reproduction represents the natural birth of species, we regard Eq.~(\ref{eq2}) as the natural birth process and will apply it to all species, including the neutral species $D$.
Eq.~(\ref{eq3}) describes the exchange of positions at rate $\varepsilon$, where $\varepsilon = 2MN$ is derived from the theory of random walks,~\cite{Redner2001A} with $N$ denoting the size of the square lattice and $M$ the mobility rate.

In addition to the fundamental interactions, we consider the natural mortality of any species at rate $f$:
\begin{equation}
	X \xrightarrow{~f~}  \varnothing \quad X \in \{A, B, C, D\}, \label{eq4}
\end{equation}
which describes the natural death of a species, leaving behind an empty site. 

\begin{figure}[ht]
	\centering
	\includegraphics[width=1\linewidth]{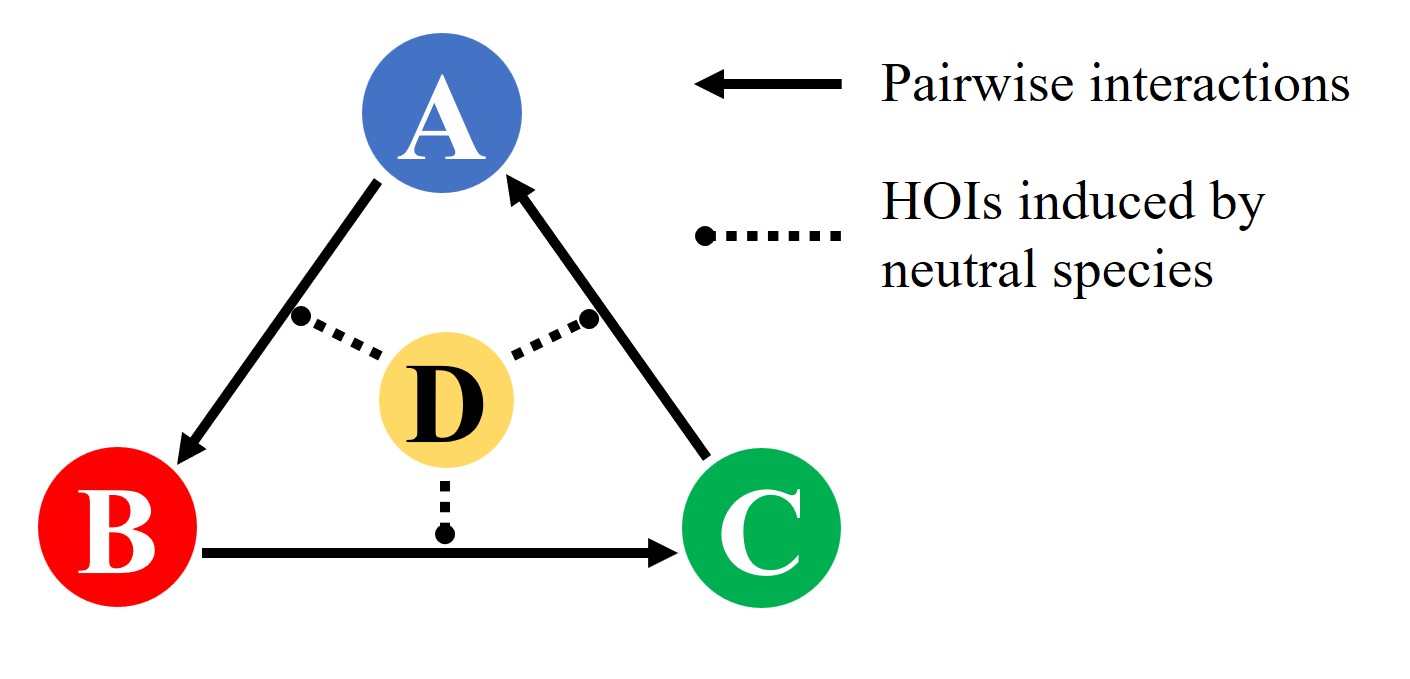}
	\caption{Species $A$, $B$ and $C$ exhibit cyclic predation following RPS rule, as indicated by the solid black arrows. Higher-order interactions induced by the neutral species $D$ significantly modulate the predation rates among the competing species, as illustrated by the dashed lines.}
	\label{fig_a}
\end{figure}

Given the influence of neutral species on higher-order processes, the intensity of competition is closely related to the proportion of species $D$. Because interactions between species -- neither predator-prey nor otherwise -- are influenced by neutral species $D$ rather than assumed to be constant, the rate $p$ for~(\ref{eq1}) can be defined as{~\cite{lu2024preUnderstanding}}:
\begin{eqnarray}
	p = \exp \left( {K \rho_i} \right), \label{eq5}
\end{eqnarray}
where $K$ represents the sensitivity coefficient for the density of species $D$ in the immediate neighborhood of node $i$, as indicated by $\rho_i$. As the intensity of intervention, the parameter $K$ can be set to the following. 
The positive $K$ indicates that the presence of species $D$ has a positive effect on interspecific competition.
Otherwise, i.e., $K \leq 0$, the role of neutral species $D$ is negative with respect to interspecific competition. 
{Pairwise interactions among competing species and higher-order interactions induced by neutral species are illustrated in Fig.~\ref{fig_a}.}

Based on the rule above, we carry out Monte Carlo simulations as follows. Initially, four species and empty sites are randomly distributed on the network. Second, select one node as the active node and one of its neighbors as the passive node. Third, the reactions described in Eqs.~(\ref{eq1})-(\ref{eq4}) occur with normalized probabilities $p/(p+q+f+\varepsilon)$, $q/(p+q+f+\varepsilon)$, $\varepsilon/(p+q+f+\varepsilon)$, and $f/(p+q+f+\varepsilon)$, respectively. One generation is defined as $N$ repetitions of the above steps. Subsequently, the second and third steps are repeated for $T$ generations, with the total simulation time set to $T = 16N$.

\section{\label{sec:result}Results}

We investigate the effect of neutral species on biodiversity from microscopic and macroscopic frameworks, with the former focusing on how neutral species affect evolutionary processes and distribution patterns, and the latter examining their effects on extinction rates.

\subsection{Characteristics of pattern formations and biodiversity}

\begin{figure}[ht]
	\centering
	\includegraphics[width=1\linewidth]{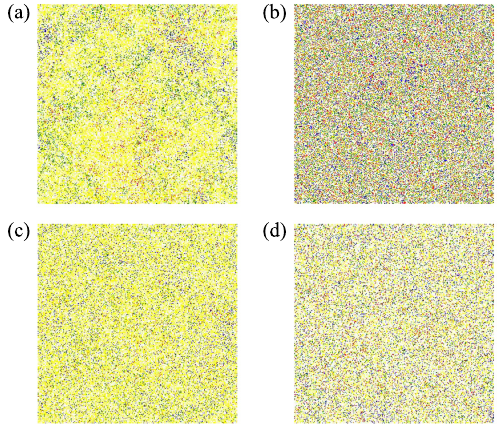}
	\caption{Characteristic snapshots under different parameter conditions of $(M, K)$:
		(a) $(M, K)=(10^{-4}, -3)$, (b) $(10^{-4}, 3)$, (c) $(10^{-3}, -3)$, and $(10^{-3}, 3)$.
		Colors in each panel, green, blue, red, yellow, and white, represent species $A$, $B$, $C$, $D$, and $\varnothing$, respectively. 
		As the sensitivity $K$ of species $D$ increases, the total density of species $A$, $B$, and $C$ increases.}
	\label{fig1}
\end{figure}

A milestone work showed that the biodiversity of three species in the spatial RPS system changes sensitively with mobility, particularly at middle and high mobility regimes, via the critical mobility $M_c = (4.5 \pm 0.5) \times 10^{-4}$.~\cite{Reichenbach2007Mobility} In addition, interspecific competition depends on the role of neutral species. To investigate microscopically how species evolve on spatially extended systems, we utilize a square lattice network of size $300 \times 300$ and consider four different parameter conditions of mobility $M$ and the intensity of intervention $K$ by species $D$: $(M, K) = (10^{-4}, -3)$, $(10^{-3}, -3)$, $(10^{-4}, 3)$, and $(10^{-3}, 3)$. Corresponding typical snapshots are presented in Fig.~\ref{fig1}, with panels organized as follows: the top and bottom correspond to the two different mobility regimes $M=10^{-4}$ and $10^{-3}$, respectively, and the left and right in each row correspond to $K=-3$ and $3$, respectively.

For $K = -3$, where the competition process is inhibited, species $D$ has a greater opportunity to expand, resulting in yellow nodes dominating the environment [see Fig.~\ref{fig1}(a)]. In addition, a higher proportion of species $D$ causes the other three species to become more dispersed. As $M$ increases from $10^{-4}$ to $10^{-3}$, the increased mobility facilitates contact between competing species, thereby slightly reducing the extent to which species $D$ inhibits competitive interactions. However, this is not sufficient to prevent species $D$ from becoming dominant. At the same time, the proportion of white nodes increases as the competitive dynamics of the cyclic competition system are suppressed. 
On the other hand, for $K = 3$, where the competition process is enhanced, the densities of species $A$, $B$, and $C$ increase significantly. Even when species other than $D$ become more densely packed, they still do not form the traditional spiral wave pattern.~\cite{Reichenbach2007Mobility} Similarly, increased mobility likely results in a higher proportion of vacant nodes, as shown in Fig.~\ref{fig1}(d).

\begin{figure}[ht]
	\centering
	\includegraphics[width=0.85\linewidth]{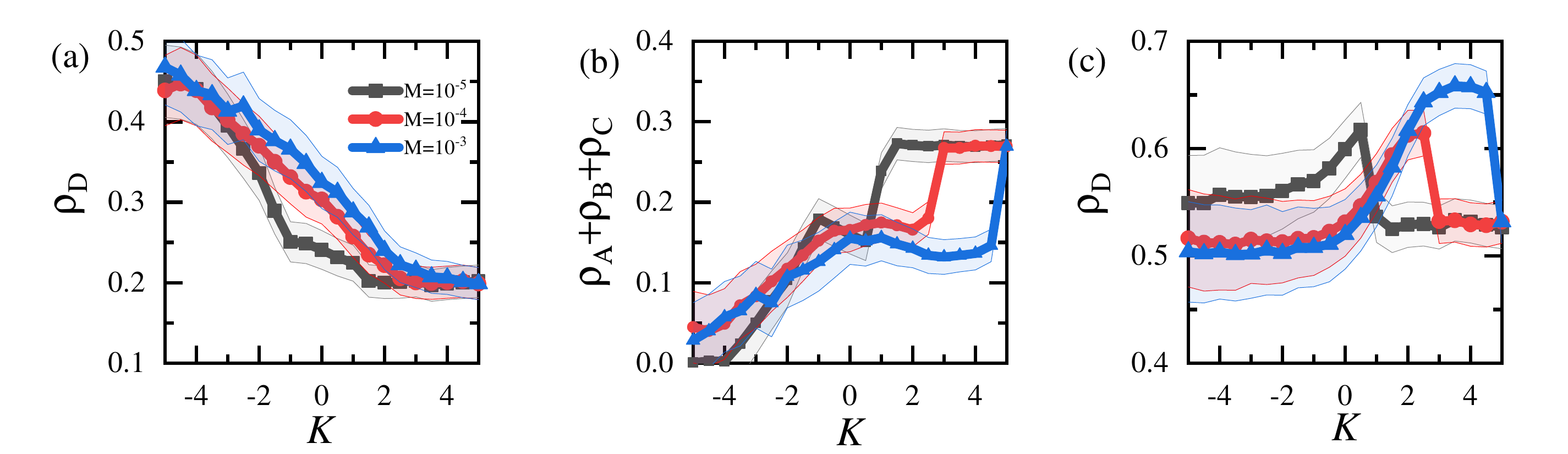}
	\caption{Synergistic effects of $K$ and $M$ on the change in the density of (a) the neutral species $D$, (b) the total density of the three species excluding $D$, and (c) the density of empty sites. 
		(a) The density of species $D$ decreases with increasing $K$ and reaches a steady state regardless of $M$. 
		(b) On the other hand, the total density of the three species can increase as $K$ increases overall. Notably, as the proportion of empty sites reaches a peak, as shown in (c), the densities decline at specific values of $K$ and $M$: $K = \pm 1$ for $M = 10^{-5}$. {The shaded region denotes the 95\% confidence interval estimated from 100 independent simulations.} For each $M$, this decline can be relaxed and reversed, leading to an increase. 
		(c) Overall, the change in empty sites shows the opposite pattern from the density of total species.} 
	\label{fig5}
\end{figure}

As shown in Fig.~\ref{fig1}, the presence of the neutral species $D$ reduces the available survival space for species $A$, $B$, and $C$. To further substantiate this conclusion, Appendix~\ref{secapp_a} provides the spatial distribution dynamics of the species under alternative parameter settings considered in the preceding sections. To investigate the effects of mobility and intervention intensity on the system state in greater depth, we examined how variations in $K$ influence species densities and vacancy density while keeping the mobility rate fixed. The results are shown in Fig.~\ref{fig5}.

More specifically, Fig.~\ref{fig5}(a) shows that the density of the neutral species $D$ decreases monotonically with increasing $K$. 
Moreover, for most values of $K$, the density levels are ordered with respect to mobility as
$M=10^{-3} > M=10^{-4} > M=10^{-5}$,
indicating that reduced mobility suppresses the persistence of species $D$. 
This mobility-induced separation becomes negligible for sufficiently large $K$ (approximately $K \gtrsim 3.5$), where the curves collapse to an essentially identical level.
 As shown in Fig.~\ref{fig5}(b), the total density of the three species $A$, $B$, and $C$ increases overall with $K$. Notably, compared with the result in Fig.~\ref{fig5}(c), when the proportion of empty sites reaches a peak, the densities decline at specific values of $K$ and $M$, e.g., $K = \pm 1$ for $M = 10^{-5}$. However, as $K$ increases further, this downward trend is eventually reversed, leading to a subsequent increase in densities. In contrast, the relationship between the fraction of empty sites and $K$ exhibits a distinct trend: as $K$ increases, the fraction of empty sites initially rises, peaks, and then declines. Furthermore, we observe that higher mobility values of $M$ require higher values of $K$ for these peaks to manifest.
 
 \onecolumngrid

\twocolumngrid

Findings on the spatially extended system showed that the total density of three species in the RPS system can increase with $K$. 
When $K$ is negative, the density of these species remains consistently low, indicating a potential threat to species coexistence. Conversely, when $K$ is positive, the density increases. This maintenance of species diversity can be attributed the factors: neutral species obstruct the competitive interactions of other species, promoting coexistence.

\subsection{Robustness of biodiversity by the extinction probability under the influence of neutral species}

\begin{figure}[ht]
	\centering
	\includegraphics[width=1\linewidth]{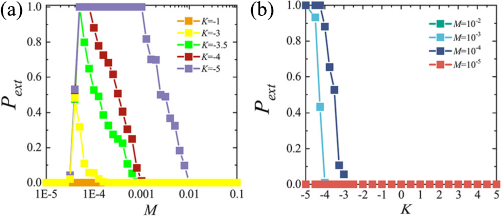}
	\caption{The extinction probability $P_{\textrm{ext}}$ as a function of: (a) $M$ for various values of $K$ and (b) $K$ for different values of $M$. In the context of higher-order interactions, increased mobility promotes coexistence, and a higher value of $K$ further enhances species coexistence.} 
	\label{fig6}
\end{figure}

When analyzing biodiversity during evolution, it is necessary to consider not only the evolutionary dynamics of species in a mean-field manner but also the extinction probability, which helps assess the robustness of biodiversity. In the microscopic framework, as $K$ increases, the higher $K$ leads to a decrease in the species density $D$, resulting in more available empty sites that foster biodiversity. 
In this regard, we present the extinction probability $P_{\text{ext}}$ for different parameter sets, as shown in Fig.~\ref{fig6}. The extinction state is considered whenever any species $A$, $B$, or $C$ becomes extinct. The extinction probability is obtained as the average over $800$ random and independent realizations.

In particular, Fig.~\ref{fig6}(a) shows how $P_{\textrm{ext}}$ varies with mobility $M$ for different values of $K$.
When $K = -1$, $P_{\textrm{ext}}$ remains at $0$ across the entire range of $M$. This result indicates that the RPS process, which involves neutral species in higher-order interactions, is favorable for species maintenance. However, this benefit comes at the expense of species populations. As $K$ decreases to $-3$, $P_{\textrm{ext}}$ initially increases, then decreases. Under these parameters, the maximum $P_{\textrm{ext}}$ reaches $0.49$, an increase compared to the previous case. 
For smaller values of $K$, such as $K = -3.5$, $-3$, and $-1$, the curve of $P_{\textrm{ext}}$ follows a similar pattern to that at $K = -3$. Different from $K=-3$, the maximum $P_{\textrm{ext}}$ increases to $1$. A lower $K$ results in a broader range of $M$ for which $P_{\textrm{ext}}$ reaches $1$. For example, $P_{\textrm{ext}} = 1$ for $M \in [10^{-4.9}, 10^{-4.7}]$ when $K = -4$, while the range for $P_{\textrm{ext}} = 1$ extends to $[10^{-4.7}, 10^{-3}]$ when $K = -5$.

On the other hand, Fig.~\ref{fig6}(b) presents $P_{\textrm{ext}}$ as a function of $K$ for various values of $M$.
For $M = 10^{-5}$ and $M=10^{-2}$, which represent excessively low and high mobility values, respectively, $P_{\textrm{ext}}$ remains $0$ regardless of $K$. Species coexistence is promoted, as observed in the RPS game with neutral species.~\cite{lu2024preUnderstanding}
For medium values of $M$, we found that $P_{\textrm{ext}}$ is consistently higher when $K$ is relatively small, specifically comparing $K = -2$ and $K = -3$. 
As $K$ increases rapidly from $-5$ to $-3$, $P_{\textrm{ext}}$ quickly converges to $0$. 
This finding suggests that, in medium-mobility regimes, low sensitivity is detrimental to coexistence, whereas high sensitivity promotes it. 
This trend is further corroborated in Fig.~\ref{fig6}(a): a smaller $K$ corresponds to a higher $P_{\textrm{ext}}$ over a moderate range of mobility.

Implementing the extinction probability shows that biodiversity can be effectively maintained at both high and low mobility levels. This pattern is also observed in rock-paper-scissors dynamics with high mortality. The extinction probability remains zero when $K > -1$. For $K < -1$, the probability initially increases to a peak before declining. Across the entire range of $K$, both extremely low and high mobility values effectively support species biodiversity. Furthermore, the natural mortality rate of species plays a crucial role in shaping biodiversity.

\subsection{Effects of a Neutral Species on System Dynamics under Asymmetric Reproduction Rates}

In real ecological systems, human interventions—such as increasing food availability or improving habitat conditions—can substantially elevate species’ reproductive rates. These enhancements often contribute to ecosystem stability and the maintenance of biodiversity. In spatial RPS models, variations in reproduction and death rates similarly generate distinctive spatial structures.

\begin{figure}[ht]
	\centering
	\includegraphics[width=1\linewidth]{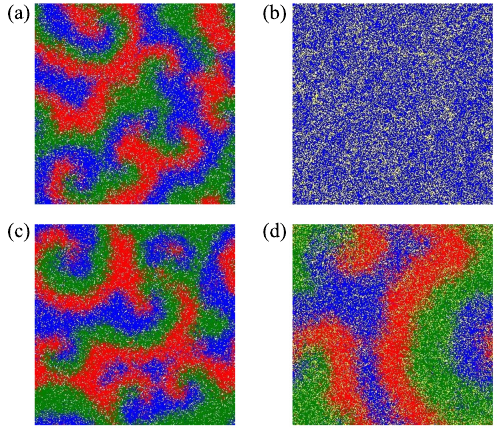}
	\caption{Characteristic snapshots of the system at high reproduction rates for different combinations of death rates for the competitive and neutral species: $(f_1, f_2)$: (a) $(1,1)$, (b) $(1,0.5)$, (c) $(0.8,1)$, and (d) $(0.8,0.5)$. The color scheme for each species is consistent with that in Fig.~\ref{fig1}. Differences in mortality between the competitive and neutral species give rise to either spiral-wave patterns or extinction of the competitive species.} 
	\label{fig9}
\end{figure}

Strong interventions from natural processes or human activities often increase species’ reproduction by several-fold; therefore, we explore a high-reproduction regime that is ecologically reasonable and relevant to both natural ecosystems and human-influenced systems. To investigate these effects, we set the reproduction rates of the competitive and neutral species to $q = 6$ and $q_N = 4$, respectively, imposing reproductive intensities much higher than the standard value $q = 1$. Here, $q_N$ denotes the reproduction rate of the neutral species. The neutral species does not directly participate in cyclic competitive interactions but coexists with the competitive species through shared space and resources. By setting $q_N = 4 < q = 6$, we introduce an asymmetry in reproductive intensity between the competitive and neutral species, reflecting realistic ecological situations in which different functional groups exhibit distinct reproductive strategies.~\cite{Szabo2007} In addition, we introduce differences in the natural death rates of the competitive and neutral species to examine how high reproduction rates, combined with various mortality configurations, influence the system’s spatial patterns. Here, $f_1$ and $f_2$ represent the mortality rates of the competitive and neutral species, respectively.
Fig.~\ref{fig9} illustrates the resulting spatial structures of the system in the steady state for the four mortality combinations: $(f_1, f_2) = (1,1), (1,0.5), (0.8,1), (0.8,0.5)$

When the death rates of all species equal the standard value of $1$, elevated reproduction rates substantially reshape the spatial pattern: the previously scattered clusters dominated by the neutral species $D$ are replaced by a spiral wave pattern dominated by the competitive species $A$, $B$, and $C$. This spatial structure markedly enhances the stability of coexistence among the competitive species, while species $D$ occupies only a very small fraction of the spatial domain. Meanwhile, the system maintains a stable, mutually reinforcing coexistence between the competitive and neutral species. When the death rate of the neutral species is reduced to $0.5$, its increased survival disrupts the stable coexistence previously supported by the competitive species and undermines the survival advantage conferred by their high reproduction rate. Figs.~\ref{fig9}(c) and~\ref{fig9}(d) illustrate the steady-state spatial patterns for $(f_1, f_2) = (0.8,1)$ and $(0.8,0.5)$, respectively. Even a modest difference in death rates is sufficient to sustain the spiral wave pattern and species coexistence, with the primary distinction being the density variations of species $D$ resulting from its different mortality levels. The temporal evolution of spatial patterns under different parameter combinations is provided in Appendix~\ref{secapp_b}.

\begin{figure}[ht]
	\centering
	\includegraphics[width=1\linewidth]{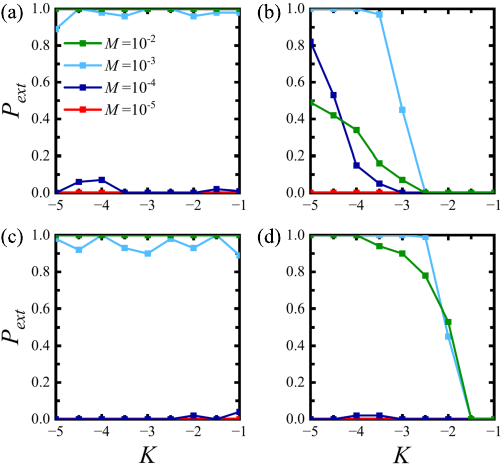}
	\caption{Extinction probabilities under high-reproduction conditions as functions of intervention intensity $K$ and mobility $M$, for different combinations of death rates for the competitive and neutral species. Panels (a)–(b) show how extinction rates vary with $K$ when the death rate of the competitive species is fixed at $f_1 = 1$ and the death rate of the neutral species is set to $1$ and $0.5$, respectively. Panels (c)–(d) present the dependence of extinction rates on $K$ for $f_1 = 0.8$ with the neutral species' death rate equal to $1$ and $0.5$.} 
	\label{fig8}
\end{figure}

\begin{figure}[ht]
	\centering
	\includegraphics[width=1\linewidth]{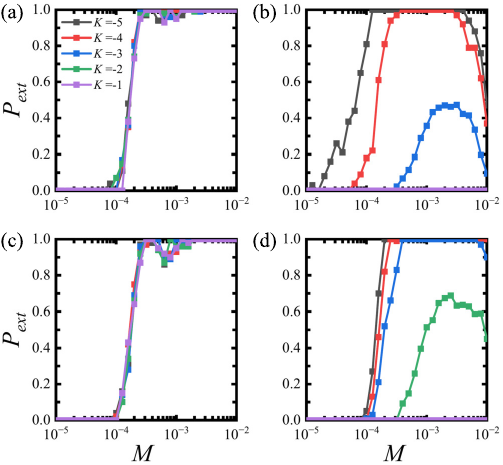}
	\caption{The variation of extinction probabilities with mobility $M$ at fixed intervention intensity $K$ across the mortality conditions $(f_1, f_2)$: (a) $(1,1)$, (b) $(1,0.5)$, (c) $(0.8,1)$, and (d) $(0.8,0.5)$.} 
	\label{fig14}
\end{figure}

High reproduction rates not only alter the system's spatial structure but also significantly influence species extinction rates. We consider four combinations of natural mortality rates $(f_1,f_2)$ for the competitive and neutral species and examine how $P_{ext}$ varies with intervention intensity $K$ and mobility $M$. The results are presented in Figs.~\ref{fig8} and~\ref{fig14}.

Fig.~\ref{fig8} shows extinction rates obtained by varying $K$ while fixing the mobility at $M = 10^{-2}$, $10^{-3}$, $10^{-4}$, and $10^{-5}$. Figs.~\ref{fig8}(a) and~\ref{fig8}(b) correspond to $f_1 = 1$ and $f_2 = 1$ and $0.5$, respectively, and reveal distinct evolutionary patterns. When $f_2 = 1$, the system exhibits only minor fluctuations in extinction rates as $K$ varies across mobility regimes. The relatively high natural mortality of the neutral species reduces its population size, thereby weakening its influence on the other species. In this regime, mobility is the dominant factor: systems with high mobility maintain consistently high extinction rates.
By contrast, when $f_2$ is reduced to $0.5$, the extinction rate decreases monotonically with increasing $K$. This indicates that a weaker suppressive effect of species $D$ on interspecific competition unexpectedly facilitates species coexistence. In other words, higher effective competitive intensity promotes the persistence of multiple species. When the competitive species death rate is reduced to $f_1 = 0.8$, as shown in (c) and (d), trends similar to those observed for $f_1 = 1$ persist.
Notably, for the parameter combination $(f_1, f_2) = (0.8, 0.5)$, a low mobility of $M = 10^{-4}$ maintains an extremely low extinction rate. This behavior contrasts with that in (b), where the dark-blue curve shows a gradual decline in $P_{ext}$ from around $0.8$ as $K$ increases. These results indicate that reduced mortality has a particularly strong positive effect on species persistence under low-mobility conditions.

Fig.~\ref{fig14} shows how extinction rates vary with mobility while the intervention intensity $K$ is held constant. The mortality parameters in each panel are set as in Fig.~\ref{fig8}, and the system exhibits similar patterns. Different mortality values for species $D$ produce qualitatively distinct behaviors, as shown in Figs.~\ref{fig14}(a,c) and Figs.~\ref{fig14}(b,d), respectively.
When the neutral species has a high mortality rate ($f_2 = 1$), $P_{ext}$ increases monotonically with mobility and ultimately approaches complete extinction, resembling the behavior observed in the standard RPS scenario. This result highlights that the inhibitory effect of species $D$ on competitive interactions requires a sufficiently large population density; when mortality is elevated, the reduced abundance of species $D$ prevents effective intervention, driving the system toward extinction.
By contrast, when $f_2$ is lower, the system exhibits a nonmonotonic response: as $M$ increases, the extinction rate rises, peaks around $M \approx 10^{-3}$, and then gradually decreases. The value of $K$ strongly influences the height of this peak; weaker intervention strengths yield lower peak extinction rates. Notably, when $K = -1$, the extinction rate remains zero across the entire mobility range, indicating that a weak intervention by the neutral species is sufficient to maintain species diversity effectively.

\begin{figure}[ht]
	\centering
	\includegraphics[width=1\linewidth]{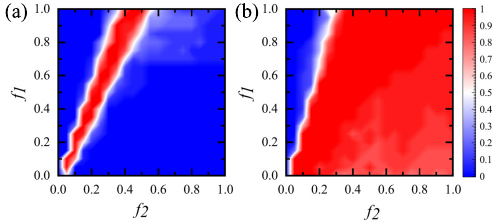}
	\caption{Heatmaps of extinction probability as a function of $f_1$ and $f_2$ for different mobility $M$: (a) $M=10^{-4}$ and (b) $10^{-3}$. Other parameters are set as follows: intervention intensity $K = -4$, competition rate $p = 1$, and reproduction rates for the competing and neutral species are $q = 6$ and $q_N = 4$, respectively. Increasing mobility markedly enlarges the extinction region in the ($f_1,f_2$) parameter space.}
	\label{fig15}
\end{figure}

To explore the synergistic effect of $f_1$ and $f_2$ on biodiversity, encompassing discussed above, we further investigate the heatmap of $P_{\rm ext}$ as a function of $f_1$ and $f_2$ for fixed mobility values, as presented in Fig.~\ref{fig15}.

As shown in Fig.~\ref{fig15}, the extinction probability exhibits a strong joint dependence on the mortality rates $f_1$ and $f_2$, and mobility further modulates this dependence. For low mobility ($M=10^{-4}$, Fig.~\ref{fig15}(a)), extinction occurs only in a narrow region of the $(f_1,f_2)$ space, emerging along a diagonal band where the mortality rates of the competitive and neutral species increase together. As $f_2$ increases further, the extinction probability decreases to near zero. In contrast, when mobility increases to $M=10^{-3}$ (Fig.~\ref{fig15}(b)), the extinction region expands substantially, with high extinction probabilities covering most of the parameter space, especially for moderate to large values of $f_1$. This comparison indicates that higher mobility amplifies the synergistic effect of species mortality rates, significantly reducing biodiversity and promoting extinction across a wider range of $(f_1,f_2)$.

\section{\label{sec:conc}Conclusion}

The mechanisms that maintain species diversity have attracted considerable attention in ecological research. Higher-order processes in species interactions are frequently observed, making their effects on biodiversity a significant focus in recent years.~\cite{Gibbs2022Coexistence} Numerous studies have established that higher-order interactions among species can facilitate coexistence.~\cite{lu2022Enhancing,lu2024preUnderstanding} In this study, we explore the impact of a fourth species, a neutral species, on species diversity within a rock-paper-scissors game. The proportion of the neutral species in the surrounding environment influences the competitive dynamics of this cyclic competition system. Furthermore, we incorporated the birth and death processes of all four species in our model.

Comprehensive modeling shows that including a fourth neutral species can facilitate the maintenance of biodiversity. Introducing this fourth species partitions resources among the three existing species, thereby reducing competitive interactions among them. In addition, in the RPS system with standard reproduction rates, the three species form a pseudo-pattern, which serves as a strategy to mitigate the potential impacts of invasion. 
By quantifying the extinction probability, we found that it approaches zero at high $K$ values, thereby corroborating macroevolutionary findings. Additionally, biodiversity is effectively sustained under high-mobility regimes.
By contrast, when reproduction rates are substantially increased, mimicking environmental changes or anthropogenic interventions, the system exhibits spiral wave patterns and, under asymmetric natural mortality, effectively alters the impacts originally induced by intervention intensity and mobility. These findings provide strong theoretical support for maintaining species coexistence by tuning differential mortality rates between competitive and neutral species.
Our study aims to clarify the importance of investigating higher-order interactions among species under the fundamental birth-death framework, ultimately providing robust evidence for the ecological significance of these interactions. We anticipate that our results will provide a sufficient basis for understanding higher-order interactions among species.

\section*{Acknowledgments}
Y.K. acknowledges support from the Yunnan Fundamental Research Projects (Grant Nos. 202501AU070193), the Scientific Research Fund of the Yunnan Provincial Department of Education (Grant No. 2025J0579) and Yunnan University of Finance and Economics (Grant No. 2025D60).
J.P. was supported by the National Research Foundation of Korea(NRF) grant 
funded by the Korea government (MSIT) (No.~RS-2023-NR076590). 
J.P. was also supported by Global-Learning \& Academic research institution for Master's$\cdot$PhD students, and Postdocs(G-LAMP) Program of the National Research Foundation of Korea(NRF) grant funded by the Ministry of Education
(No.~RS-2025-25442355).	
This work was also supported by the Scientific Research Fund Project of Yunnan Provincial Education Department (No.~2025Y0804).

\appendix

\section{\label{secapp_a}Spatial pattern evolution of the system under different combinations of neutral-species mobility and intervention intensity}

\begin{figure}[ht]
	\centering
	\includegraphics[width=1\linewidth]{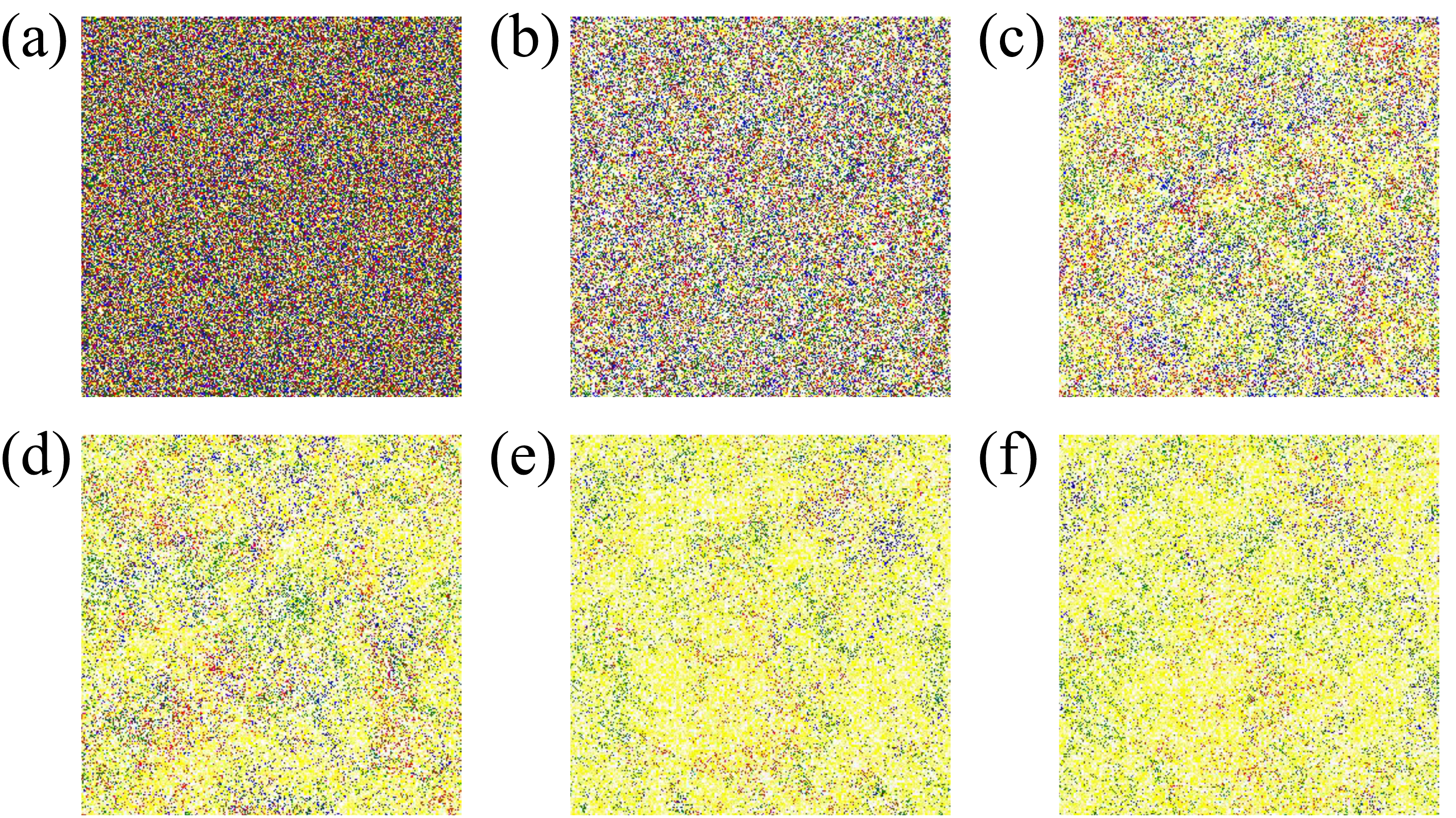}
	\caption{Typical snapshots for fixed parameters $(K,M)=(-3,10^{-4})$ at different time steps $t$: (a) $1$, (b) $4000$, (c) $600000$, (d) $700000$, (e) $800000$, (f) $900000$. Color indications are the same as Fig.~\ref{fig1}. Competing species form pseudo-clusters, which help maintain species coexistence.}
	\label{figS1}
\end{figure}

To investigate how intermediate mobility affects species evolution, we present evolutionary snapshots with parameters $K = -3$ and $M = 10^{-4}$ as illustrated in Fig.~\ref{figS1}. 

Fig.~\ref{figS1}(a) shows the initial distribution of the system, with the four species and empty spaces randomly distributed. As time progresses, more empty spaces are created, facilitating reproduction [see Figs.~\ref{figS1}(b-c)]. {In a system with low mobility, competition among species can increase species density by reducing the likelihood of interactions between different species.}

\begin{figure}[ht]
	\centering
	\includegraphics[width=1\linewidth]{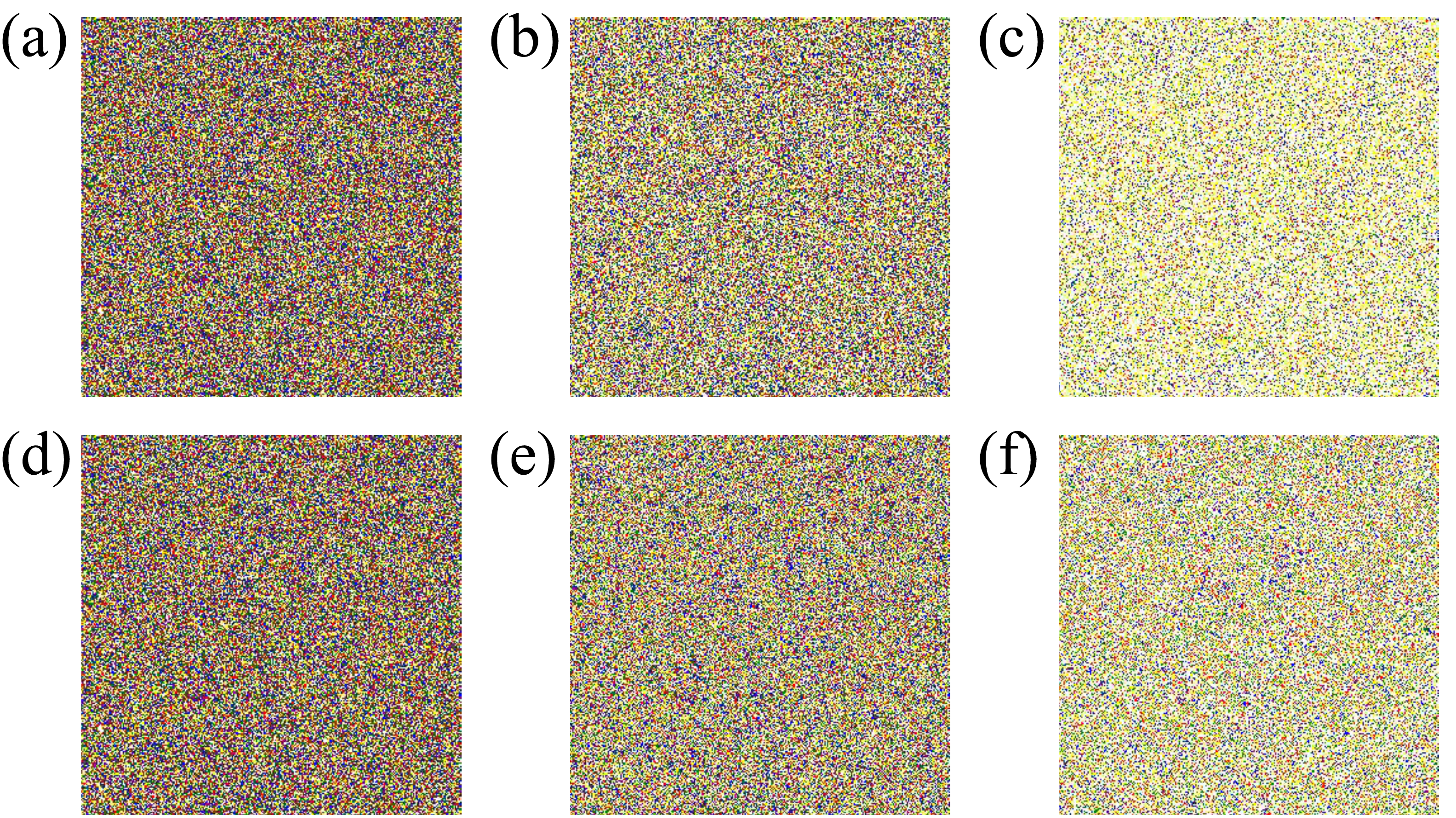}
	\caption{Typical snapshots for fixed $K=-3$ with different $M$: (a-c) $M = 10^{-3}$ and (d-f) $M=10^{-4}$. Different time steps are considered in each row: $1$, $4000$, and $900,000$ from the left to the right. Higher sensitivity coefficients are conducive to increasing the cumulative densities of species excluding species $D$. Color indications are the same as above.}
	\label{figS3}
\end{figure}

To illustrate species' evolution under the high-intensity intervention of neutral species, we present evolutionary snapshots at different mobility values, as shown in Fig.~\ref{figS3}: $M = 10^{-3}$ for tops and $M = 10^{-4}$ for bottoms. Specifically, when $M = 10^{-3}$, as shown in Figs.~\ref{figS3}(a-c), high mobility increases the likelihood of encounters between different species. The presence of species $D$ further intensifies competition among these species, leading to a continuous increase in the number of white nodes as the system transitions from the initial state [see Fig.~\ref{figS3}(a)] to the steady state. Consequently, there is no clear single dominance of species $A$, $B$, $C$, or $D$, and empty spaces persist in the steady state. As the mobility decreases to $M = 10^{-4}$, the system transitions from the initial distribution [Fig.~\ref{figS3}(d)] to a steady state [Fig.~\ref{figS3}(f)]. At this point, the likelihood of encounters among competing species decreases, leading to an increase in the proportion of each species.

In summary, comparing evolutionary snapshots with those under system stabilization, we find that species mobility significantly affects competitive interactions. High species mobility reduces species densities by increasing contact among competitive species pairs. When the intervention intensity of species $D$ is hostile, species $D$ becomes dominant, and medium mobility rates contribute to the emergence of pseudo-clusters within the system. In contrast, high mobility rates disrupt these clusters, resulting in a more diffuse separation among species $A$, $B$, and $C$. Conversely, when the intervention intensity is positive, the four species cannot form a dominant species; although they do not form clusters, they can still stabilize their existence.

\section{\label{secapp_b}Evolution of spatial structures under high reproduction rates with asymmetric mortality}

{High reproduction rates cause substantial changes in the system’s spatial structure. Here, we present two typical spatial patterns and compare how spatial structures evolve over time under asymmetric mortality rates.}

\begin{figure}[ht]
	\centering
	\includegraphics[width=1\linewidth]{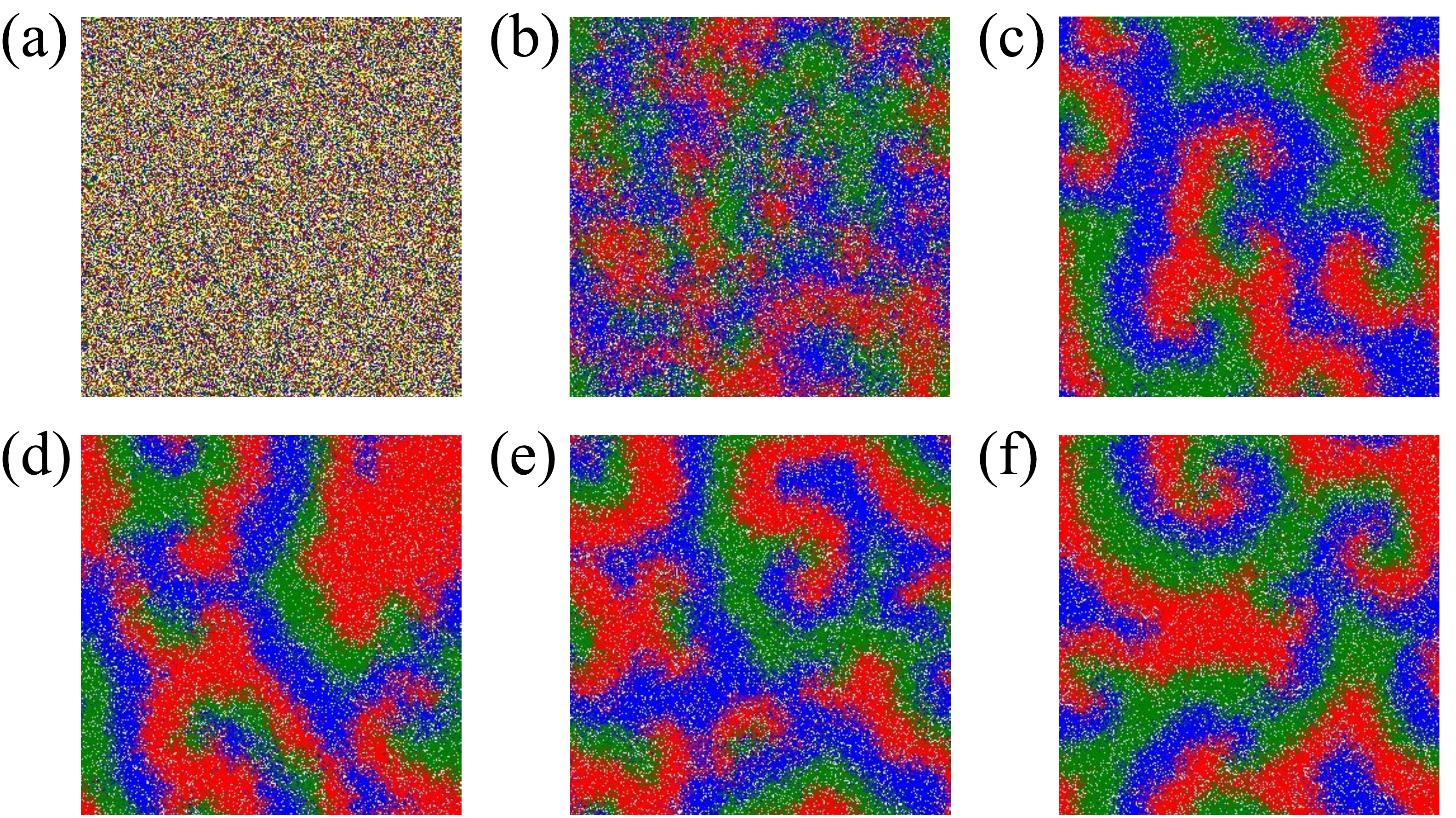}
	\caption{Snapshots of the system dynamics under the standard mortality setting with fixed parameters $(K,M)=(-4,10^{-4})$ at different time steps $t$: (a) $1$, (b) $300$, (c) $30000$, (d) $150000$, (e) $300000$, and (f) $450000$. The species represented by each color follow the same conventions as in earlier figures. The competitive species form spiral wave pattern that greatly enhance the stability of coexistence among them, while the neutral species persists at a relatively small but stable proportion of the spatial domain.}
	\label{figS4}
\end{figure}

By setting the death rates of the competitive and neutral species to $(f_1, f_2) = (1,1)$ and fixing $(K,M)=(-4,10^{-4})$, we examined the temporal evolution of the system’s spatial patterns at various time steps, as shown in Fig.~\ref{figS4}.

Fig.~\ref{figS4} shows the spatial patterns at different time steps. The high reproduction rate drives the system to form a spiral wave pattern in which the competitive species reach a stable equilibrium, while species $D$ occupies only a small fraction of the space yet remains persistently present. The elevated reproductive intensity enhances both system stability and species diversity, thereby promoting coexistence between the competitive and neutral species to a greater extent.

\begin{figure}[ht]
	\centering
	\includegraphics[width=1\linewidth]{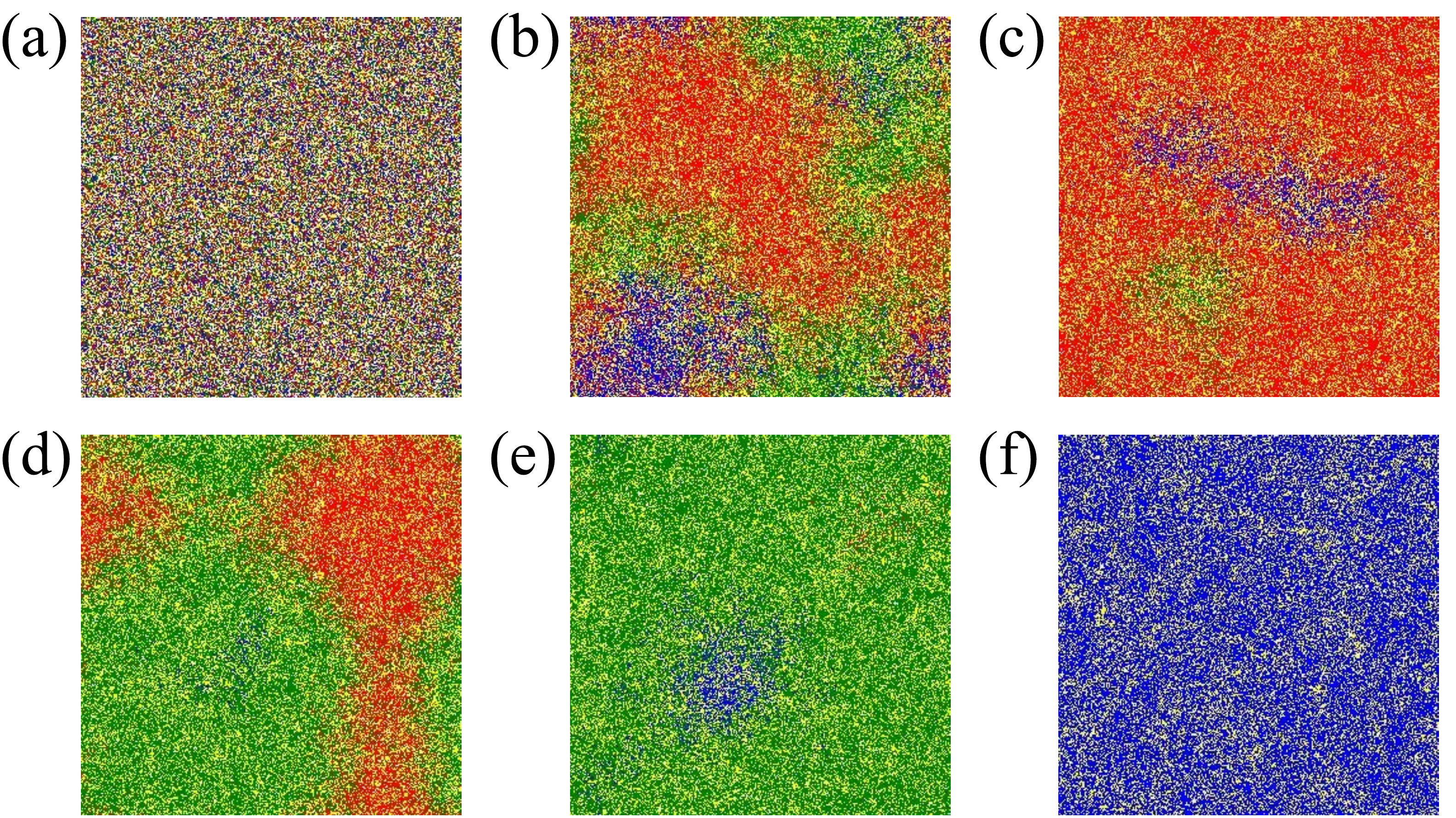}
	\caption{Temporal evolution of the spatial patterns when the competitive species retain the standard death rate while the mortality of species $D$ is reduced to $f_2 = 0.5$. The species exhibit pronounced spatial aggregation, and the system eventually evolves into a state dominated by species $B$. The values of $K$, $M$, and the time steps $t$ in panels (a–f) are the same as those in Fig.~\ref{figS4}.}
	\label{figS5}
\end{figure}

When the mortality of species $D$ is reduced to $f_2 = 0.5$, the system exhibits behavior distinct from the standard mortality setting, as shown in Fig.~\ref{figS5}. At steady state, species $A$ and $C$ go extinct, while species $B$ dominates the system and establishes stable coexistence with species $D$. The reduced mortality enhances the survival prospects of species $D$ but simultaneously disrupts the coexistence of all species.

\begin{thebibliography}{99}
\bibitem{Cardinale2002Species}
B. J. Cardinale, M. A. Palmer, and S. L. Collins,
``Species diversity enhances ecosystem functioning through interspecific facilitation,"
Nature {\bf 415}, 426--429 (2002). 
% https://doi.org/10.1038/415426a

\bibitem{Schmidt2011Persistence}
M. W. I. Schmidt, M. S. Torn, S. Abiven, T. Dittmar, G. Guggenberger, I. A. Janssens, M. Kleber, I. K{\"o}gel-Knabner, J. Lehmann, 
D. A. C. Manning, P. Nannipieri, D. P. Rasse, S. Weiner, and S. E. Trumbore,
``Persistence of soil organic matter as an ecosystem property,"
Nature {\bf 478}, 49--56 (2011).
% https://doi.org/10.1038/nature10386

\bibitem{Brophy2017Biodiversity}
C. Brophy, {\'A}. Dooley, L. Kriwan, J. A. Finn, J. McDonnell, T. Bell, W. M. Cadotte, and J. Connolly,
``Biodiversity and ecosystem function: making sense of numerous species interactions in multi-species communities,"
Ecology {\bf 98}, 1771--1778 (2017).
% https://doi.org/10.1002/ecy.1872

\bibitem{Reichenbach2007Mobility}
T. Reichenbach, M. Mobilia, and E. Frey,
``Mobility promotes and jeopardizes biodiversity in rock–paper–scissors games,"
Nature {\bf 448}, 1046--1049 (2007).
% https://doi.org/10.1038/nature06095

\bibitem{Park2017Emergence}
J. Park, Y. Do, B. Jang, and Y.-C. Lai,
``Emergence of unusual coexistence states in cyclic game systems,"
Sci. Rep. {\bf 7}, 7465 (2017).
% https://doi.org/10.1038/s41598-017-07911-4

\bibitem{szolnoki2024When}
A. Szolnoki and X. Chen,
``When faster rotation is harmful: The competition of alliances with inner blocking mechanism,"
Phys. Rev. Research {\bf 6}, 023087 (2024).
% https://doi.org/10.1103/PhysRevResearch.6.023087

\bibitem{szolnoki2016zealots}
A. Szolnoki and M. Perc,
``Zealots tame oscillations in the spatial rock-paper-scissors game,"
Phys. Rev. E {\bf 93}, 062307 (2016).
% https://doi.org/10.1103/PhysRevE.93.062307

\bibitem{szolnoki2004Phase}
A. Szolnoki and G. Szab{\'o},
``Phase transitions for rock-scissors-paper game on different networks,"
Phys. Rev. E {\bf 70}, 037102 (2004).
% https://doi.org/10.1103/PhysRevE.70.037102

\bibitem{Bairey2016High}
E. Bairey, E. D. Kelsic, and R. Kishony,
``High-order species interactions shape ecosystem diversity,"
Nat. Commun. {\bf 7}, 12285 (2016).
% https://doi.org/10.1038/ncomms12285

\bibitem{Majhi2022Dynamics}
S. Majhi, M. Perc, and D. Ghosh,
``Dynamics on higher-order networks: A review,"
J. R. Soc. Interface {\bf 19}, 20220043 (2022).
% https://doi.org/10.1098/rsif.2022.0043

\bibitem{Majhi2017Dynamics}
F. Battiston, G. Cencetti, I. Iacopini, V. Latora, M. Locas, A. Patania, J.-G. Young, and G. Petri,
``Networks beyond pairwise interactions: Structure and dynamics,"
Phys. Rep. {\bf 874}, 1--92 (2017).
% https://doi.org/10.1016/j.physrep.2020.05.004

\bibitem{szolnoki2020pattern}
A. Szolnoki, B.F. de Oliveira, and D. Bazeia,
``Pattern formations driven by cyclic interactions: A brief review of recent developments,"
EPL {\bf 131}, 68001 (2020).
% 10.1209/0295-5075/131/68001

\bibitem{szolnoki2020strategy}
A. Szolnoki and X. Chen,
``Strategy dependent learning activity in cyclic dominant systems,"
Chaos Soliton. Fract. {\bf 138}, 109935 (2020).
% https://doi.org/10.1016/j.chaos.2020.109935

\bibitem{sinervo1996rock}
B. Sinervo and C. M. Lively,
``The rock–paper–scissors game and the evolution of alternative male strategies,"
Nature {\bf 380}, 240--243 (1996).
% https://doi.org/10.1038/380240a0

\bibitem{kirkup2004antibiotic}
B. C. Kirkup and M. A. Riley,
``Antibiotic-mediated antagonism leads to a bacterial game of rock--paper--scissors in vivo,"
Nature {\bf 428}, 412--414 (2004).
% https://doi.org/10.1038/nature02429

\bibitem{neumann2010evolutionary}
G. F. Neumann and G. Jetschke,
``Evolutionary classification of toxin mediated interactions in microorganisms,"
BioSystems {\bf 99}, 155--166 (2010).
% https://doi.org/10.1016/j.biosystems.2009.10.007

\bibitem{nahum2011evolution}
J. R. Nahum, B. N. Harding, and B. Kerr,
``Evolution of restraint in a structured rock-paper-scissors community,"
Proc. Natl. Acad. Sci. U.S.A. {\bf 108}, 10831--10838 (2011).
% https://doi.org/10.1073/pnas.1100296108

\bibitem{duan2024does}
X. Duan, J. Ye, Y. Lu, C. Du, B. Jang, and J. Park,
``Does cooperation among conspecifics facilitate the coexistence of species?"
Chaos Soliton. Fract. {\bf 186}, 115308 (2024).
% https://doi.org/10.1016/j.chaos.2024.115308

\bibitem{lu2022Effects}
Y. Lu, X. Wang, M. Wu, L. Shi, and J. Park,
``Effects of species vigilance on coexistence in evolutionary dynamics of spatial rock–paper–scissors game,"
Chaos {\bf 32}, 093116 (2022).
% https://doi.org/10.1063/5.0103247

\bibitem{Souza2017Apex}
C. A. Souza-Filho, D. Bazeia, and J. Ramos,
``Apex predator and the cyclic competition in a rock-paper-scissors game of three species,"
Phys. Rev. E {\bf 95}, 062411 (2017).
% https://doi.org/10.1103/PhysRevE.95.062411

\bibitem{lu2022Enhancing}
Y. Lu, C. Shen, M. Wu, C. Du, L. Shi, and J. Park,
``Enhancing coexistence of mobile species in the cyclic competition system by wildlife refuge,"
Chaos {\bf 32}, 081104 (2022).
% https://doi.org/10.1063/5.0093342

\bibitem{Avelino2019Predominance}
P. P. Avelino, B. F. de Oliveira, and R. S. Trintin,
``Predominance of the weakest species in Lotka-Volterra and May-Leonard formulations of the rock-paper-scissors model,"
Phys. Rev. E {\bf 100}, 042209 (2019).
% https://doi.org/10.1103/PhysRevE.100.042209

\bibitem{Liao2020Survival}
M. J. Liao, A. Miano, C. B. Nguyen, L. Chao, and J. Hasty,
``Survival of the weakest in non-transitive asymmetric interactions among strains of {\it E. coli},"
Nat. Commun. {\bf 11}, 6055 (2020).
% https://doi.org/10.1038/s41467-020-19963-8

\bibitem{lu2024preUnderstanding}
Y. Lu, X. Wang, C. Du, Y. Wang, Y. Geng, L. Shi, and J. Park,
``Understanding the role of neutral species by means of high-order interaction in the rock-paper-scissors dynamics,"
Phys. Rev. E {\bf 109}, 014313 (2024).
% https://doi.org/10.1103/PhysRevE.109.014313

\bibitem{Chatterjee2022Controlling}
S. Chatterjee, S. N. Chowdhury, D. Ghosh, and C. Hens,
``Controlling species densities in structurally perturbed intransitive cycles with higher-order interactions,"
Chaos {\bf 32}, 103122 (2022).
% https://doi.org/10.1063/5.0102599

\bibitem{Gibbs2022Coexistence}
T. Gibbs, S. A. Levin, and J. M. Levine,
``Coexistence in diverse communities with higher-order interactions,"
Proc. Natl. Acad. Sci. U.S.A. {\bf 119}, e2205063119 (2022).
% https://doi.org/10.1073/pnas.220506311

\bibitem{Lu2025Enhancement}
Y. Lu, H. Dai, H. Tan, X. Duan, L. Shi, and J. Park,
``Enhancement of persistence in the rock-paper-scissors dynamics through higher-order interactions,"
Appl. Math. Comput. {\bf 487}, 129083 (2025).
% https://doi.org/10.1016/j.amc.2024.129083

\bibitem{2007SzabSegregation}
Szabó G, Szolnoki A, Sznaider G A.,
``Segregation process and phase transition in cyclic predator-prey models with an even number of species,"
Phys. Rev. E {\bf 76(5)}, 051921 (2007).


\bibitem{2019BazeiaInvasion}
Bazeia D, De Oliveira B F, Szolnoki A.,
``Invasion-controlled pattern formation in a generalized multispecies predator-prey system,"
Phys. Rev. E {\bf 99(5)}, 052408 (2019).

\bibitem{2018SzolnokiEvolutionary}
Szolnoki A, Perc M.,
``Evolutionary dynamics of cooperation in neutral populations,"
New Journal of Physics {\bf 20(1)}, 013031 (2018).

\bibitem{2021LiBeyond}
Li Y, Mayfield M M, Wang B, et al. Beyond direct neighbourhood effects: higher-order interactions improve modelling and predicting tree survival and growth[J]. National Science Review, 2021, 8(5): nwaa244.

\bibitem{PDHL:2013}
J. Park, Y. Do, Z.-G. Huang, and Y.-C. Lai,
``Persistent coexistence of cyclically competing species in spatially extended ecosystems,"
Chaos {\bf 23}, 023128 (2013).
% https://doi.org/10.1063/1.4811298

\bibitem{Park:2018}
J. Park,
``Asymmetric interplay leads to robust coexistence by means of a global attractor in the spatial dynamics of cyclic competition,"
Chaos {\bf 28}, 081103 (2018).
% https://doi.org/10.1063/1.5048468

\bibitem{PJ:2021}
J. Park and B. Jang, 
``Structural stability of coexistence in evolutionary dynamics of cyclic competition,"
Appl. Math. Comput. {\bf 394}, 125794 (2021).
% https://doi.org/10.1016/j.amc.2020.125794

\bibitem{PJ:2023}
J. Park and B. Jang, 
``Role of adaptive intraspecific competition on collective behavior in the rock-paper-scissors game,"
Chaos Soliton. Fract. {\bf 171}, 113448 (2023).
% https://doi.org/10.1016/j.chaos.2023.113448

\bibitem{Xiao:2000}
Y. Xiao,
``A general theory of fish stock assessment models,"
Ecol. Modell. {\bf 128}, 165--180 (2000).
% https://doi.org/10.1016/S0304-3800(00)00199-X

\bibitem{LiChen:2009}
X. Li and Y. Chen,
``Age structure, growth and mortality estimates of an endemic {\it Ptychobarbus dipogon} (Regan, 1905) (Cyprinidae: Schizothoracinae) in the Lhasa River, Tibet,"
Environ. Biol. Fishes {\bf 86}, 97--105 (2009).
% https://doi.org/10.1007/s10641-008-9371-5

\bibitem{Holt:2013}
C. Jørgensen and R. E. Holt,
``Natural mortality: Its ecology, how it shapes fish life histories, and why it may be increased by fishing,"
J. Sea Res. {\bf 75}, 8 --18 (2013).
% https://doi.org/10.1016/j.seares.2012.04.003

\bibitem{Then:2014}
A. Y. Then, J. M. Hoenig, N. G. Hall, and D. A. Hewitt,
``Evaluating the predictive performance of empirical estimators of natural mortality rate using information on over 200 fish species,"
ICES J. Mar. Sci. {\bf 72}, 82--92 (2014).
% https://doi.org/10.1093/icesjms/fsu136

\bibitem{McCoy:2008}
M. W. McCoy and J. F. Gillooly,
``Predicting natural mortality rates of plants and animals,"
Ecol. Lett. {\bf 11}, 710--716 (2008).
% https://doi.org/10.1111/j.1461-0248.2008.01190.x

\bibitem{Knebel2015coupleddeath}
J. Knebel, M. F. Weber, T. Kr{\"u}ger, and E. Frey E,
``Evolutionary games of condensates in coupled birth–death processes,"
Nat. Commun. {\bf 6}, 6977 (2015).
% https://doi.org/10.1038/ncomms7977

\bibitem{Yu2016QBD}
Q. Yu, D. Fang, X. Zhang, C. Jin, and Q. Ren,
``Stochastic evolutionary dynamic of the rock-scissors-paper game based on a quasi birth and death process,"
Sci. Rep. {\bf 6}, 28585 (2016).
% https://doi.org/10.1038/srep28585

\bibitem{West2020fixation}
R. West and M. Mobilia,
``Fixation properties of rock-paper-scissors games in fluctuating populations,"
J. Theor. Biol. {\bf 491}, 110135 (2020).
% https://doi.org/10.1016/j.jtbi.2019.110135

\bibitem{He2012Coexistence}
F. He, D. Y. Zhang, and K. Lin,
``Coexistence of nearly neutral species,"
J. Plant. Ecol. {\bf 5}, 72--81 (2012).
% https://doi.org/10.1093/jpe/rtr040

\bibitem{Redner2001A}
S. Redner, {\it A guide to first-passage processes} (Cambridge University Press, 2001).

\bibitem{Szabo2007}
G. Szabó and G. Fath,
``Evolutionary games on graphs," 
Phys. Rep. {\bf 446}, 97--216 (2007).
% https://doi.org/10.1016/j.physrep.2007.04.004
\end{thebibliography}
\end{document}